\documentclass[aps,twocolumn,showkeys]{revtex4}

\pdfoutput=1
                             
\usepackage{graphicx}                                       
\usepackage{amssymb}
\usepackage{wasysym}
\usepackage{caption}
\usepackage{lipsum}
\usepackage[a4paper,top=15mm,headheight=50mm,%
   bottom=15mm,left=12mm,right=12mm]{geometry}

\newcommand{\umafigura}[3]{ 
  \noindent%
  \begin{minipage}{\linewidth}
  \makebox[\linewidth]{
    \includegraphics[width=\linewidth]{#1}}
   \captionof{figure}{#3}\label{#2}
  \end{minipage}
}

\newcommand{\eqt}[2]{
    \begin{equation}
             {#2}
             \label{#1}
    \end{equation}
}

\begin{document}

%
%
%
\title{Bootstrap Calibration of Inductive Voltage Dividers at Inmetro}
\author{
  F. A. Silveira${}^{}$\footnote{Email address: {\it fsilveira@inmetro.gov.br}} 
}

\affiliation{
  Instituto Nacional de Metrologia, Qualidade e Tecnologia,\\
  Avenida N. S. das Gra\c cas 50, 25250-020 D. Caxias RJ, Brazil}

\date{\today}

\begin{abstract}

The Laboratory of Metrology in Electrical Standards of Inmetro (Lampe) has built,
in 2010, a transformer which is able to perform inductive voltage divider ratio
calibrations using the bootstrap method. This transformer was idealized according 
to the design adopted at {\it Physikalisch-Technische Bundesanstalt} (PTB), and has 
very similiar characteristics to that system.
Lampe's bootstrap system is not yet in operation, and DIT calibrations are
presently being carried according to a triangular set of measurements involving standard
capacitors. This method, though, has uncertanties limited to a few parts in $10^6$; the
primary objective of Lampe's bootstrap system project is to achieve uncertanties 
at least 10 times smaller than this.

\end{abstract}

\keywords{
Measurement, electrical standards, inductive voltage dividers, bootstrap
}

\maketitle

%
%
%
\section{Introduction}
\label{intro}

The project of a DIT calibration system at Lampe
is centered on a two-stage transformer capable of making bootstrap measurements
of the ratio between inductive voltage dividers (IVD) steps, as thoroughly described 
in \cite{kibble,hall1968}.
Inmetro's bootstrap transformer has been constructed 
in a 2010 cooperation with 
PTB \cite{kyriazis2012} (and is itself based on the PTB system), in order to develop 
a system capable of calibrating the 10:-1 ratio of the main IVD used
in coaxial 2- and 4-terminal-type impedance bridges. 

Lampe has two capacitance bridges which have been in operation for 5 to 6 years now.
These bridges are an important 
link in the traceability chain to derive the capacitance unit from the
quantum Hall resistance \cite{schurr2002}.
And with the intent of showing how we plan to improve our calibration uncertainty 
of measurements carried in these bridges, this work presents a brief description
of our bootstrap system 
project, and points a few important characteristics of the 
bootstrap transformer constructed here.

%
%
%

\section{System Diagram}
\label{diagrams}

Following the original PTB project, the bootstrap transformer constructed at Inmetro is
capable of providing voltage (both in-phase and quadrature) to two current injection 
systems simultaneously.
Fig.~\ref{bootstrapsys} shows an updated schematics for the complete IVD calibration setup; 
in this figure, we highlight the bootstrap transformer constructed at Inmetro
\cite{kyriazis2012}, at the extreme left of this drawing.
This transformer is two-stage-type, and is apt to calibrate two-stage
IVDs at 10:-1 ratio. For a revision of the two-stage principle of transformers, 
as well as the bootstrap method, the reader is referred to \cite{kibble,hall1968}.

The potencial difference to be used as reference to the taps of the object IVD is provided through
the triaxial connectors {\tt FH,FL}. These two output terminals, and
both IVD and bootstrap windings wired by them, make the main loop of the calibration circuit.
The potencial difference between any given IVD taps $n,n+1$ for each $n=-1,\cdots, 9$ 
is to be compared to the the reference voltage $U_{Ref}$ at the bootstrap terminals 
{\tt FH} and {\tt FL}, and is denoted \cite{kibble}
 \eqt{diferenca}{U_n-U_{n-1}=U_{Ref}\left(1+\alpha_{n}\right),}
where $\alpha_n$ is the reading of the ratio on the knobs of the panel of the injection inductive 
decade $CN_1$. $CN_1$ is a 6-decade,
single-core coaxial divider with two output taps; $T_1$ is a 100:1 ratio toroidal transformer.
The subsystem formed by $CN_1$, $T_1$ and the RC phase shifter make the voltage injection system 
over the main loop, to be described in Sec. \ref{injection}.

In the triaxial branches, the outermost shields of the cables are taken to guard potencial 
levels, which equal the potencials of the two IVD taps to be measured. These guard 
potencials are selected through two cursors (labelled $SW_1$ in Fig.~\ref{bootstrapsys}) 
that run jointly between the potencials of ${\tt A_{10},A_{-1}}$.
Guarding is meant to supress the parasitic currents from the center conductors of
$\tt FH,FL$ terminals by taking them to the same potencial as the innermost shield of the cables.
For a revision of the guarding/shielding principles, see, e.g., \cite{morrison}

The bridge is balanced by connecting two taps of the IVD to the $\tt FH,FL$ terminals of
the bootstrap transformer. Then a guard potencial should be adjusted on $SW_2$ that
suits to the selected taps of the IVD. Finally, $CN_1$ and $CN_2$ are serially
adjusted. All these steps are sequentially and iteractivelly repeated until the 
voltage read through  the lock-in amplifier fluctuates below 30 
or 20 $\mu$V, when the net gain in the pre- and lock-in amplifier stages are set 
to about $10^3$.

\newlength{\myboxwidth}
\addtolength{\myboxwidth}{2\columnwidth}
\addtolength{\myboxwidth}{\columnsep}
\newsavebox{\mybox}
\savebox{\mybox}{\begin{minipage}{\myboxwidth}
   \captionof{figure}{
   Electrical schematics for Inmetro's bootstrap transformer \cite{kyriazis2012}. 
   The potencial difference to be used as reference to the taps of the object IVD is provided 
   through the triaxial connectors {\tt FH,FL}, and $\tt SW_1$ selects the guard potencial 
   applied to the outermost shield of the cables.
   The isolation transformer is provided with a wagner-type compensation 
   subcircuit, labelled $CN_2$.
   The bootstrap transformer injects voltage into the main loop through $\tt CN_1$,
   $\tt T_1$ and a phase shifter, which provide in-phase and quadrature components over 
   the {\tt FH} terminal. 
   A similar system $\tt T2$ then detects the total voltage difference on the {\tt FL} terminal 
   and read it through low noise pre- and lock-in amplifiers. 
   The cables drawn in orage are choked, in order to minimize the effects of electromagnetic
   noise over closed loops of the circuit.
   }\label{bootstrapsys}
 \end{minipage}}
\begin{figure*}[!t]
   \centering
   \includegraphics[width=\linewidth]{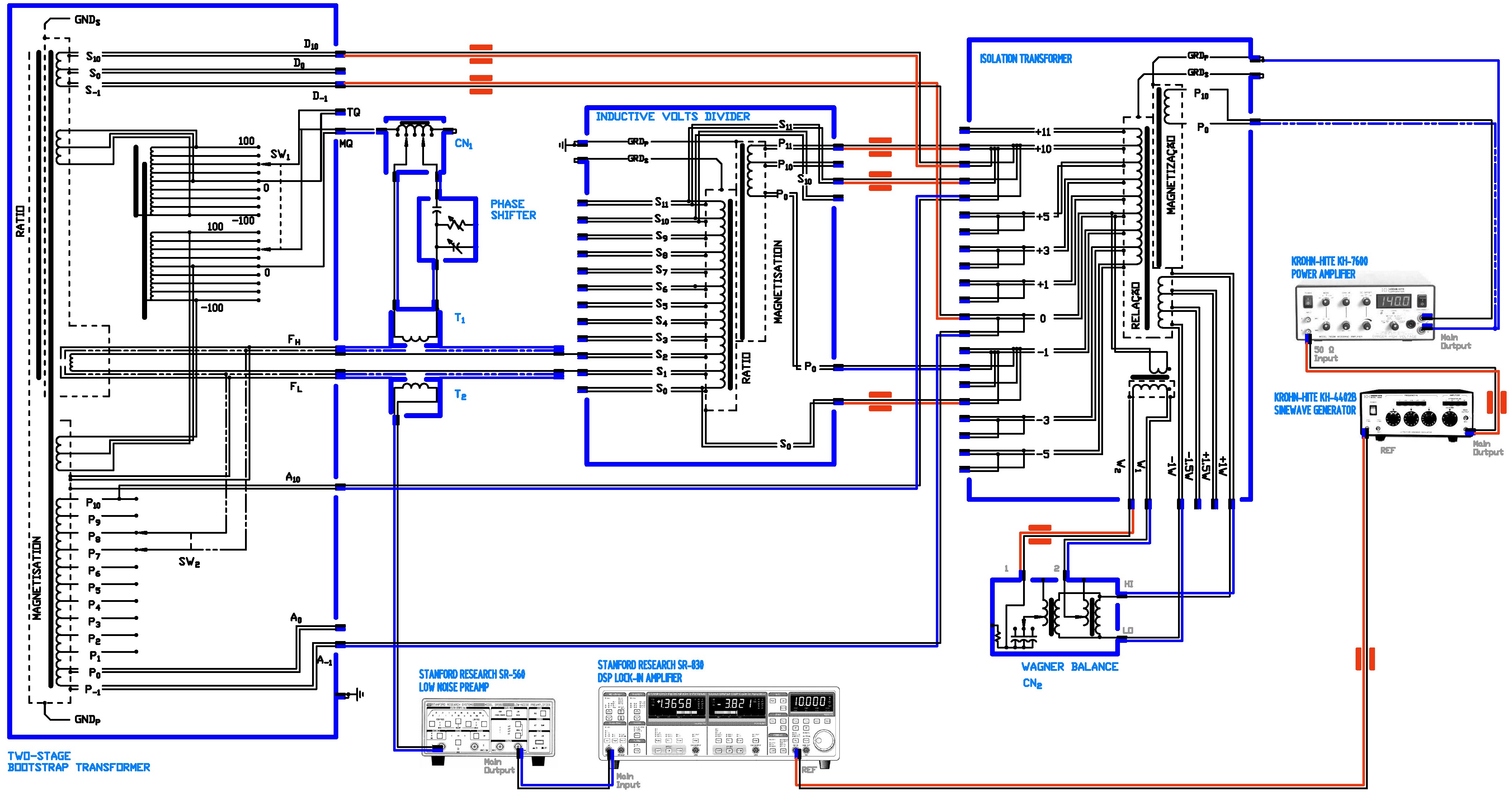}
   \usebox{\mybox}
\end{figure*}

%
%
%

\subsection{Supply and compensation}
\label{supply}

The system shown in Fig.~\ref{bootstrapsys} is supplied through a sinewave generator, 
a power amplifier and an isolation two-stage transformer, which provide the needed 
voltage submultiples. 
The isolation transformer output provide the proper supply to the IVD to be calibrated 
(shown at the center of Fig.~\ref{bootstrapsys}) and the bootstrap transformer 
(shown at the extreme left of the schematics.)

The transformer is supplied through coaxial terminals ${\tt A_{10},A_{-1}}$ (magnetising stage) and
$\tt D_{10},D_{-1}$ (ratio stage.) Connectors labelled {\tt TQ,MQ} supply in-phase and quadrature 
components to (the magnetising stage of) the injection system.
In practice, however, the inductive decades available in Lampe to build the injection system 
are single-core type. So, only the output {\tt MQ} of the transformer is connected to the current 
injection system. 

IVD-based bridge comparison circuits estabilish the relation between the IVD ratio in one arm
and unknown impedances on the other, under the constraint that equilibrium holds between two
defined nodes. Parasitic admittances on the (coaxial) ports of the circuit elements cause
current deviations that may be important to the final balance. 

In order to compensate for the effect
of these parasitics, we couple between the source terminals
and ground a RC complex circuit 
that balances the outer shield potencial of the cables to the ground potencial, thus
minimising the stray currents at the ports of the circuit elements.
The most widely used source compensation circuit on impedance bridges of the kind 
discussed here is the {\it wagner circuit}
\cite{kibble,hague}. The isolation transformer is provided with a wagner-type compensation 
subcircuit, all-contained in the chassis of the inductive decade
labelled $CN_2$ in Fig.~\ref{bootstrapsys}.

%
%
%

\subsection{Injection/detection}
\label{injection}

The injection decade $CN_1$ is combined to a phase shifter (made out of an RC network) at the output
which is meant to provide a small quadrature component. Both in-phase and quadrature components
of $CN_1$ full voltage are then mixed in the injection 100:1 transformer coupled to the
inner conductor of {\tt FH}, and
the whole system add up to inject small, adjustable 
in-phase and quadrature voltage components over the main loop.
The impedances of the phase shifter whirl about 100 $\Omega$ (trimmed resistance) and 1 pF
(adjustable capacitance).
A similar system detects the total voltage difference on the {\tt FL} terminal 
and read it through low noise pre- and lock-in amplifiers, where the null condition of the
bridge is read in both its in-phase and quadrature components.

The potencials of the pairs of terminals {\tt MP,MQ} and {\tt TP,TQ} are  selected 
through individual switches to ratios $\pm 0,0001$ to $\pm 1$ of the potencial difference 
between the terminals {\tt FH} and {\tt FL}. 
These potencials are straight derivations from the  magnetising ({\tt MP,MQ}) and
 ratio ({\tt TP,TQ}) windings, and can provide voltage source to the magnetising and ratio stages of
a two-stage combining network. For the sake of concision, only $\tt TQ$ and $\tt MQ$ are 
shown in Fig.~\ref{bootstrapsys}. The omitted $\tt TP,MP$ derivations, coupled to the magnetising
and ratio windings, are in everything symmetric to $\tt TQ,MQ$, though.

%
%
%

\section{Injection transformer detail}
\label{detail}

Aside $\tt FH,FL$ triax terminals, the connections to the bootstrap and the 
injection/detection transformers all are of the coaxial BPO 
type; the $\tt FH,FL$ terminals are of the 1051 DKE/Fischer type.
Fig \ref{exploded} shows an exploded view of such an injection/detection transformer.
It shows the coaxial chassis with BPO- and Fischer-type connectors, the toroidal 
windings and the central triaxial cable that works as the one-winding coil of this
device.

The guarding at the point where $\tt T_1$ (alternativelly, $\tt T_2$) couples to 
$\tt FH$ ($\tt FL$) brings about some delicate management of the internal shielding 
of this transformer.
A special geometry of the two shield layers of the cable is required to avoid stray 
currents and the resulting leakage inductance.

Fig.~\ref{internview} shows a plane section of the windings coupling. At the 
middle of the triax cable shown in this figure, there is a small section of the
shields (not explicit shown in the picture) that leaves the center conductor electrically
exposed to the magnetic flux in the toroidal core.
This sectioning follows the tecniques described in \cite{kibble} for minimising
stray capacitances between the shields sections.
These parasitic effects are 
well discussed in \cite{kibble} and references therein, and they can be
avoided (or at least greatly attenuated) with a particular, distinct superposition of 
the shields at the point of magnetic coupling also applied in \cite{kyriazis2012}.
Its application to $\tt T_1, T_2$,
as well as its effectiveness will be detailed elsewhere, after further tests have been
carried on.

\vskip .5cm
\umafigura{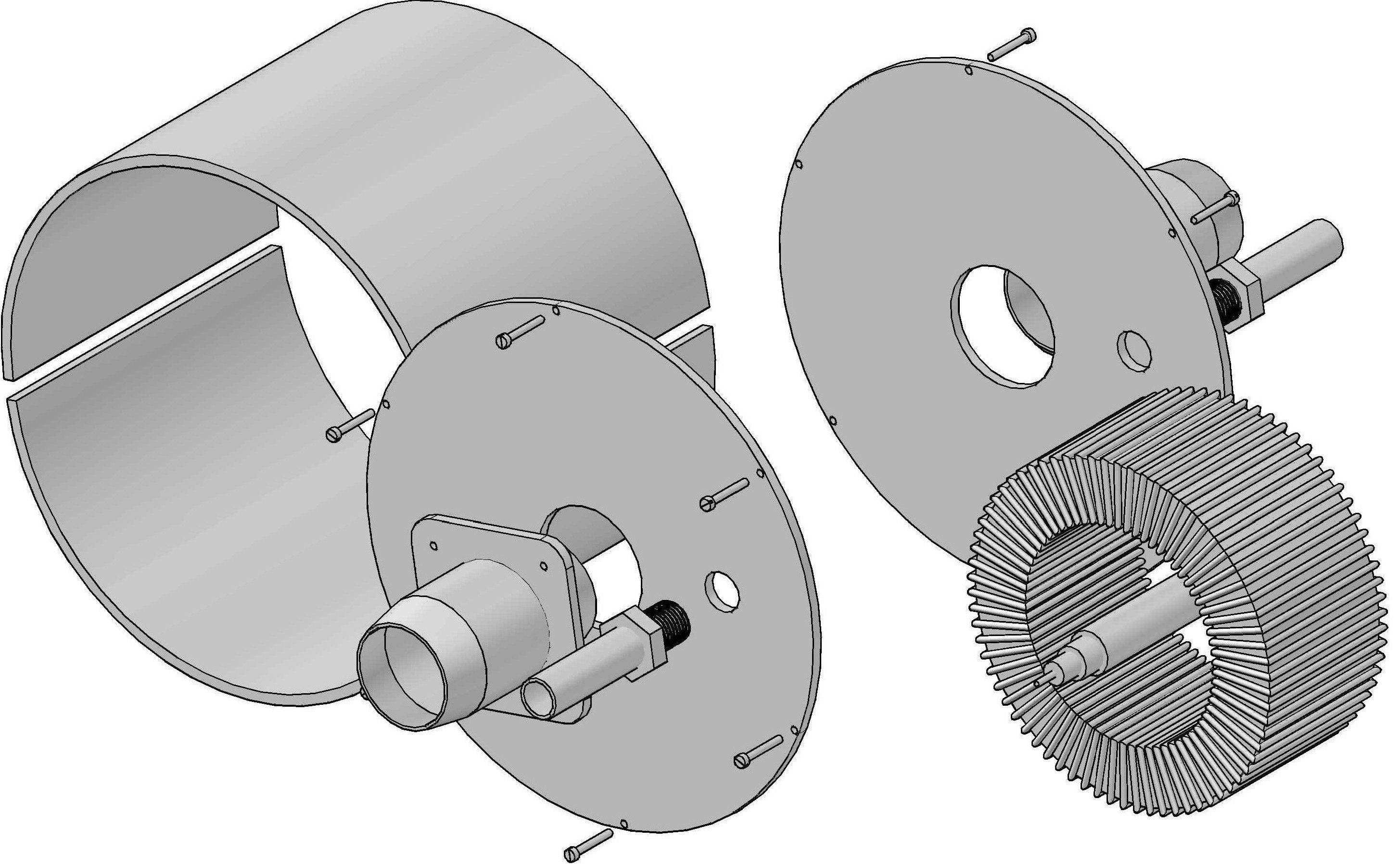}{exploded}{The diagram shows the exploded view of the 
coaxial chassis, with its BPO- and Fischer-type connectors, the toroidal 
windings and the central triaxial cable that works as the one-winding coil of this
device.}

\vskip .5cm
\umafigura{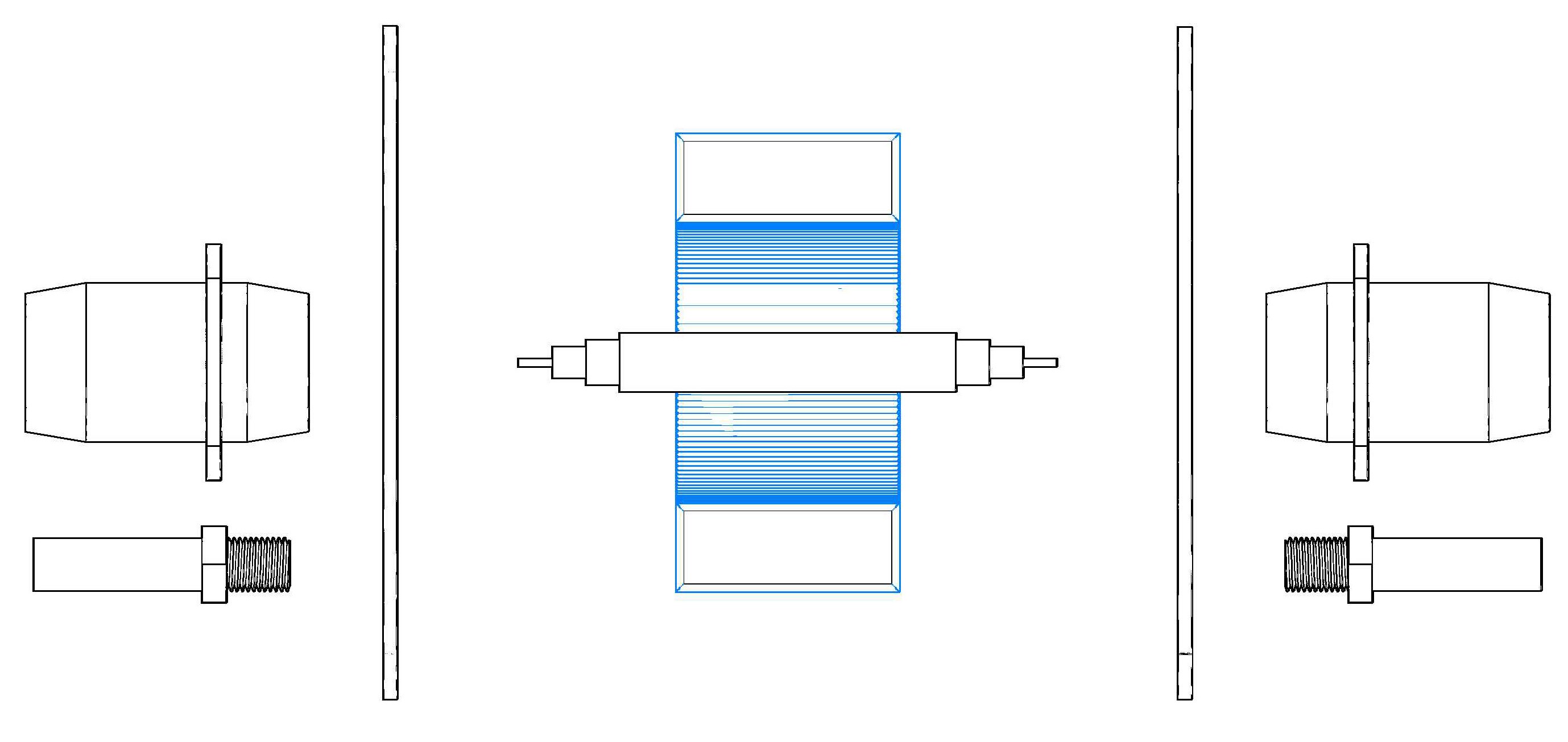}{internview}{Plane section of the windings coupling
of $\tt T_1$ and $\tt T_2$. The middle of the triax cable has a small section in the
shields (not explicit shown) that couples it to the windings on the toroidal core.}

%
%
%

\section{Conclusions}
\label{conclusao}

As said before,
Lampe's bootstrap measurement system is not yet in operation, and DIT calibrations are
presently being carried based on the external calibration values of standard
capacitors. This method  has uncertanties limited to a few parts in $10^6$, which may be 
considered to be large for this type of measurement. In addition,
having our standards calibrated externally involves large costs, 
both on financial and logistic sides.

The main objective of Lampe's bootstrap system project is to achieve uncertanties 
in the $10^{-7}$ or $10^{-8}$ range.
The circuit element on which the system is based, the bootstrap transformer was built in 2010.
This project went into a halt for some time now, and measurements still must be made on 
the transformer to test for some of its characteristics, like its behaviour under 
loading conditions and stability.

We're still short of but a few circuit elements in order to assemble the bootstrap bridge
system shown in Fig \ref{bootstrapsys}, but the next stages are already in progress.
We have already submitted to Inmetro's precision workshop the projects of Figs.~\ref{exploded}
and~\ref{internview}, which should be ready soon,
and started the building of cabling and peripheral connections.
For instance, we already dispose of a rich supply of choke cores, 
an important part of any AC precision coaxial circuit used to equalize
currents between inner and outer shields of the cables, and thus minimise the
effects of electromagnetic noise on closed loops \cite{kibble,morrison,hague}.

Also, most of the connectors and cabling are of the coaxial BNC/RG-56, 50$\Omega$ commercial 
type, and are not currently a problem. All in all, we expect to start making measurements
over the next year.

\section{Acknowledgements}
\label{ack}

The author thanks G. A. Kyriazis for the support in retrieving notes on
the Inmetro-PTB cooperation, as well as for
important suggestions on the design of the injection/detection systems.

%
%
%

\onecolumngrid\ \vfill\twocolumngrid

\end{document}